# The origin of the density scaling exponent for polyatomic molecules and the estimation of its value from the liquid structure

F. Kaśkosz[1,*], K. Koperwas[1,*], A. Grzybowski[1] and M. Paluch[1]

1. University of Silesia in Katowice, Institute of Physics, 75 Pułku Piechoty 1, 41-500 Chorzów, Poland

* corresponding author: filip.kaskosz@us.edu.pl, kajetan.koperwas@us.edu.pl

## ABSTRACT

In this article, we unravel the problem of interpreting the density scaling exponent for the polyatomic molecules representing the real van der Waals liquids. Our studies show that the density scaling exponent is a weighted average of the exponents of the repulsive terms of all interatomic interactions occurring between molecules, wherein the potential energy of a given interaction represents its weight. It implies that potential energy is a key quantity required to calculate the density scaling exponent value for real molecules. Finally, we use the well-known method for potential energy estimation and show that the density scaling exponent could be successfully predicted from the liquid structure for fair representatives of the real systems.

**MANUSCRIPT**

Undoubtedly, the density scaling law evidently differs from the numerous conceptions describing how the system dynamics depends on the thermodynamic conditions. The uniqueness of this idea results not only from the fact that it expresses quite complex thermodynamics of the supercooled liquid as a function of only one variable but also from the fact that it makes it using exclusively one parameter. Interestingly, this material constant (the density scaling exponent, $\gamma$) is directly related to the intermolecular interaction potential which makes the density scaling law delivers an insight into the effective microscopic interactions occurring between molecules. The latter is especially intriguing in the context of the studies on real materials because the hardly-to-be-obtained information on the intermolecular interactions can be straightforwardly deduced from the easily accessible macroscopic properties. However, a definition of $\gamma$ for polyatomic molecules that actually mimic the real molecules, is needed.

When the density scaling is observed, the dynamical property of the real liquid, $X$, e.g., the structural relaxation time, diffusion constant, or viscosity, is identical for all thermodynamic conditions characterized by the same value of the scaling argument, $Tv^\gamma$ (where $T$ is the temperature and $v = V/N$ is the systems' volume per molecule, i.e., the volume of the system $V$ divided by the number of the systems' molecules $N$). Consequently, $X$ values collapse onto one master curve:

$$X = \mathcal{F}(Tv^\gamma) \qquad\qquad Eq.~(1)$$

However, the scaling property predicted by Eq. ( 1) is derived only for simple model systems consisting of the soft spheres [1,2], which interact via the entirely symmetric intermolecular potential having the form of Inverse Power Law ($IPL$),

$$U_{IPL}(r) = Cr^{-m} + A \qquad\qquad Eq.~(2)$$

where $A$ represents the attractive background, $C$ are potential parameters, whilst $m$ is the exponent of the repulsive term directly related to the density scaling exponent,

$$\gamma = m/3. \qquad\qquad Eq.~(3)$$

This scaling property of the $IPL$ systems was initially known as their quasi-universality [3–8], and it still attracts a significant scientific interest. It is due to the fact that the $IPL$ is recognized as a fair approximation of the Lennard-Jones (LJ) potential [9,10], which is the theoretically justified for the real liquids, and also because the density scaling has been reported not only for the simple LJ systems [11–14] but for more realistic model liquids too. [13,15–18] The theoretical background for those findings is delivered by the virial $\left(W(r) = -\frac{1}{3}r\frac{dU(r)}{dr}\right)$ and potential energy correlation [12,13], the proportionality constant of which is identified as $\gamma$. [19–22] This correlation constitutes the framework of the isomorph theory and the concept of Roskilde-simple (R-simple) liquids. [23] The phase diagrams of R-simple liquids exhibit isomorphs, i.e., curves having the constant invariant excess entropy. These curves link thermodynamic states at which the particle distribution functions, normalized time-autocorrelation functions, and transport coefficients are identical when expressed in so-called reduced units. It also must be noted that the performed computational experiments demonstrate that, generally, $\gamma$ is not constant but depends on the thermodynamic conditions.

However, the isomorphs are not found for polyatomic model systems with flexible bonds. Nevertheless, an empirical scaling of the dynamics of those systems is still possible, and therefore the mentioned curves are called 'pseudoisomorphs'. [24] The latter are expected to be observed in real materials, for which the first report on the density scaling was delivered by Tölle, who analyzed the quasi-elastic neutron scattering data for canonical van der Walls liquid ortho-terphenyl and pointed out that the observed dynamic crossover could be characterized by an effective constant value resulted in the scaling variable $Tv^4$. [25,26] Next, Dreyfus successfully scaled rotational relaxation times, obtained from light-scattering data for different isotherms of ortho-terphenyl, onto a single master curve as a function of $Tv^4$. [27] The numerous subsequent attempts to scale the dynamics of various glass formers revealed that the exponent of 4 could not be treated as a universal value for all systems. [28–36] Nevertheless, the latter does not change the fact that the density scaling of the molecular dynamics can be achieved for more than one hundred glass formers [37] even at a wide range of pressures [38,39], when $\gamma$ between 2 and 7 is used. The all-mentioned experimental studies have led to the emergence of fundamental questions on the physical factors influencing $\gamma$ and the way to predict its value for real systems, i.e., for systems in which molecules are constructed from many interacting centers. Intuitively, the diversity of $\gamma$ values might be ascribed to the differences in some effective intermolecular interaction potential, which depends on the kind and arrangement of atoms creating the molecules. However, till now, the exact definition of $\gamma$ for polyatomic molecules has not been proposed. It makes that desired prescription for $\gamma$ estimation using some macroscopic characteristic of the system, ideally determined at a single thermodynamic state, is far from being achieved.

In this article, basing on the simulations of the molecular dynamics of quasi-real model systems we give an exact definition of $\gamma$ for polyatomic systems, which atoms interact by purely repulsive potentials. A key finding of our research is the proposed method for $\gamma$ calculation, which uses the system's structure at a single thermodynamic state. Thereby, we provide solid grounds for explaining the origin of density scaling exponent value for real materials.

The system we examine is comprised of the quasi-real model molecules, the atoms of which are arranged in a rhombus shape (rhombus-like molecule, RM). Our choice is justified by the fact that RM mimics the typical features of real molecules, i.e., structural anisotropy and flexibility, while their simplicity limits the number of factors impeding an understanding of the obtained results. [40] The crucial is also that the RMs, which are consisted of 4 identical atoms interacting according to *IPL* with $m = 12$, obey the density scaling law with $\gamma = 4$. [41] Hence, Eq. ( 3) is fulfilled, which implies that an RM system studied in Ref. [41] is perfectly suitable for the planned studies. In this system, the atom-atom interactions were set on the basis of the LJ potential parameters established by the OPLS all-atom force field [42] for the carbon atoms of the benzene ring, i.e., we took the corresponding $\varepsilon, \sigma$ and defined $C = 4\varepsilon\sigma^{12}$ ($\varepsilon$ characterizes a 'depth' of the LJ potential, whereas $\sigma$ sets a position of its minimum). To introduce the second type of atoms, we redefine $m$ for 2 atoms lying alongside the shorter diagonal of the RM. The atoms of the new type interact by *IPL* with $m = 18$, whereas the mixed interactions are parameterized by $m = 15$. As a consequence, the used potentials are as follows: $U_{AA}(r) = C_{AA}r^{-12}$, $U_{BB}(r) = C_{BB}r^{-18}$, and $U_{AB}(r) = C_{AB}r^{-15}$, where $C_{AA} = 4\varepsilon\sigma^{12}, C_{BB} = 4\varepsilon\sigma^{18}$ and $C_{AB} = 4\varepsilon\sigma^{15}$. For all interactions, the attractive constant is set to assure that potential energy is equal to 0 at the distance where the intermolecular interactions are cut-off, $r_{cut} = 1.12\ nm$. The consistency between inter- and intra-molecular interactions is preserved since the stiffness of the molecule is set using parameters derived by the OPLS force field for the benzene ring.

In the first step of our studies, we performed a set of computer simulations of molecular dynamics using GROMACS software. [59,60] The system, which is comprised of the 2048 RMs, was cooled down at isobaric conditions equal to 50, 80, 160, 300, 500, and 1000 MPa with the decrement of the temperature equal to 5 K, using Martyna-Tuckerman-Tobias-Klein barostat [64,65] and Nose-Hoover [61–63] thermostat. Each simulation run lasts for 10 million steps (with timestep $dt = 0.001\ ps$). The first half of the simulation run was devoted to the system equilibration, whereas the average system volume was estimated from the data collected through the second half of the simulation. Then we performed the *NVT* simulations (the isomorph theory is valid at this ensemble), which lasted long enough to observe the entire relaxation of the incoherent intermediate scattering function (IISF) [43] calculated for the centers of the molecule's mass. This procedure enabled estimation of the system relaxation time, $\tau$, which was subsequently reduced according to $\tau^* = \frac{\tau}{v^{\frac{1}{3}}\sqrt{m_{mol}/(k_B T)}}$, where $k_B$ is Boltzmann constant and $m_{mol}$ is the molecule mass. [21] The results are presented in Fig. 1a.

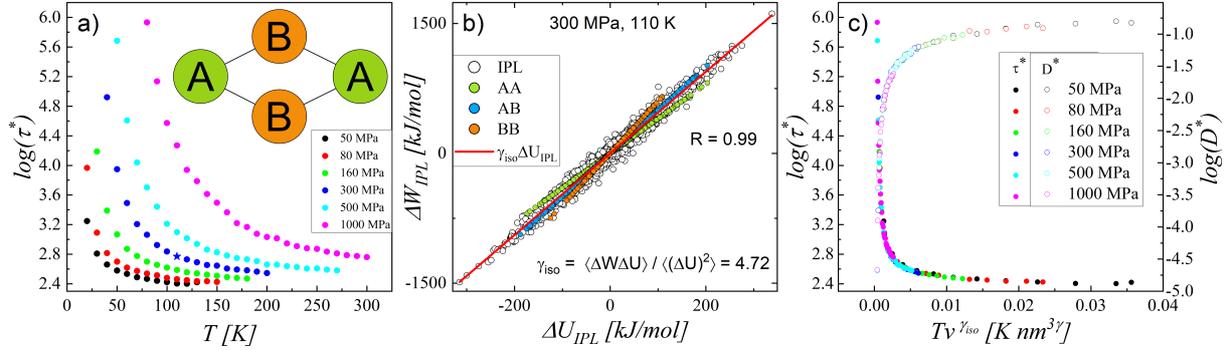

*Fig.1*
*Panel a) presents the reduced relaxation time determined at studied thermodynamic conditions covering a wide range of pressure and temperatures. The scheme of the RM is shown as an inset. Panel b) depicts the fluctuations of the virial and the potential energy resulting from interatomic interactions for all particular non-bonded interactions and their sum. The resulting slope of the dependence is equal to 4.72, which has been determined from a single thermodynamic state, T=110K and P=3000MPa. The R parameter represents a correlation coefficient value. Panel c) represents the density scaling of reduced relaxation times and diffusion constants.*

Following the isomorph theory, the density scaling exponent could be determined from the single thermodynamic state because it describes the correlation between fluctuations of total virial and total potential energy, $\Delta W = \gamma \Delta U$, where $\Delta$ denotes an instantaneous value of a given quantity minus its average value. However, it has to be recalled that for systems with intramolecular interactions this correlation is not fulfilled. Nevertheless, the correlation is still observed for the components of virial and potential energy, which originate from the non-bonded interactions [41]. Consequently, the density scaling exponent can be determined from the following equation:

$$\gamma_{iso} = \frac{\langle \Delta W_{IPL} \Delta U_{IPL} \rangle}{\langle (\Delta U_{IPL})^2 \rangle}, \qquad Eq.~(4)$$

where $\langle \rangle$ means constant-volume canonical averages. [12] In Fig. 1b, we examine a single thermodynamic state, which is characterized by $T = 110\ K$ and $p = 300\ MPa$, and as one can observe, $\Delta W_{IPL}$ and $\Delta U_{IPL}$ are almost perfectly correlated with slope equal to $\gamma_{iso} = 4.72$. This value enables accurate scaling of $\tau^*$ and also the reduced diffusion constants, $D^* = D\sqrt{\frac{m}{k_BT}}/v^{1/3}$ ($D$ is determined from the mean square displacement), see Fig. 1c. At this point, it has to be mentioned that consistently with the isomorph theory, $\gamma_{iso}$ differs in various thermodynamic conditions. However, its variation is slight for the studied system, and it does not significantly influence the quality of the density scaling ($\gamma_{iso}$ varies between 4.63 at $T = 20\ K$ and $p = 80\ MPa$ and 4.80 at $T = 300\ K$ and $p = 1000\ MPa$).

Since the examined herein RM system obeys the density scaling law, we could focus on elucidating the origin $\gamma_{iso}$ value. The RM is characterized by 3 different types of interactions, $U_{AA}(r), U_{AB}(r), U_{BB}(r)$, which are described by $IPL$ with $m$ equal to 12, 15, and 18 and none of those values leads directly to $\gamma_{iso} = 4.72$. However, from Fig. 1b, we know that the linear relationship $\Delta W_{IPL} = \gamma \Delta U_{IPL}$ is valid at any timestep $i$ thus $W_{IPL,i} - \overline{W}_{IPL} = \gamma(U_{IPL,i} - \overline{U}_{IPL})$.

Calculating the virial $W(r) = -\frac{1}{3}r\frac{dU(r)}{dr}$, we get $W_{IPL,i} + \bar{R} = \gamma U_{IPL,i}$, where $\bar{R}$ is an intercept, which results from vanishing the interaction potential constant term $A$. Taking into account that within the studied system one can distinguish in general 4 different interaction we can reorganize the left side of the above equation: $W_{AA,i} + W_{AB,i} + W_{BA,i} + W_{BB,i} + \bar{R} = \gamma U_{IPL,i}$, where $W_{AA,i}$ is the virial resulting from interactions between atoms of type $A$ at time $i$. Because $W_{AA}$ is a linear function of corresponding potential energy, $U_{AA}$, with the slope described by $m$ parameter of given interaction potential, we can take advantage from $W_{AA} = \gamma_{AA}U_{AA} - R_{AA}$, which leads to $(\gamma_{AA}U_{AA,i} - R_{AA,i}) + (\gamma_{AB}U_{AB,i} - R_{AB,i}) + (\gamma_{BA}U_{BA,i} - R_{BA,i}) + (\gamma_{BB}U_{BB,i} - R_{BB,i}) + \bar{R} = \gamma(U_{AA,i} + U_{AB,i} + U_{BA,i} + U_{BB,i})$ and then, summing up times we get $\gamma = \frac{\gamma_{AA}\Sigma_i(U_{AA,i}) + \gamma_{AB}\Sigma_i(U_{AB,i}) + \gamma_{BA}\Sigma_i(U_{BA,i}) + \gamma_{BB}\Sigma_i(U_{BB,i}) + \Sigma_i(\bar{R} - R_{AA,i} - R_{AB,i} - R_{BA,i} - R_{BB,i})}{\Sigma_i(U_{AA,i} + U_{AB,i} + U_{BA,i} + U_{BB,i})}$, where the whole term, including intercepts $R$, vanishes.

Interestingly, the sums of the potential energies can be immediately replaced by the ensemble averages:

$$\gamma = \frac{\gamma_{AA}\langle U_{AA}\rangle + \gamma_{AB}\langle U_{AB}\rangle + \gamma_{BA}\langle U_{BA}\rangle + \gamma_{BB}\langle U_{BB}\rangle}{\langle U_{AA}\rangle + \langle U_{AB}\rangle + \langle U_{BA}\rangle + \langle U_{BB}\rangle}. \qquad Eq.\ (5)$$

Eq. ( 5) is a weighted average of the exponents describing particular intermolecular interactions, with weights determined by average potential energies originating from a corresponding interactions. Hence, the latter is an exact interpretation of the $\gamma$ for many-atomic molecular systems described by IPL potentials.

The form of Eq. ( 5) leads to two important consequences. The first one is that if all interatomic interactions are described by the *IPL* with the same value of $m$, i.e., all the differences between interactions are included in various $C$ (see Eq. ( 2)), the straightforwardly resulted $\gamma$ should be equal to $m/3$. Naturally, in this case the average potential energies $\langle U_{XX}\rangle$, *XX* represents a type of interactions, differ from each other, but all of them implies the same value of $\gamma$. We test this prediction at the end of the paper. The second consequence, which in our opinion is more crucial from the point of view of the real molecules, refers to an example of the molecules constructed from at least two atoms. Consider two kinds of molecules created from the same atoms but differently arranged. In such a case, the interatomic potentials for given atom-atom interactions are identical, predefined, and do not depend on the molecular structure. However, different molecular structures might imply that the same atom-atom interactions are mostly registered at different distances. Consequently, $\langle U_{XX}\rangle$ for interactions are expected to be different, which immediately implies different $\gamma$ values. Hence, the determination $\langle U_{XX}\rangle$ is a key to determine $\gamma$ for molecules composed of many different atoms. Ideally, the latter would be done using a macroscopic characteristic of the system, which is relatively easier to obtain than a microscopic one. We propose to employ the well-known pair-correlation function, $g(r)$. This function characterizes the structure of the system and will be different for discussed situation of molecules of identical atomic composition. Moreover, $g(r)$ depends on the thermodynamic conditions. Thus, it naturally implies the reported variation in $\gamma$ while the temperature and density are changed.

In the case of RM studied herein, $g(r)$ can be calculated for 4 different pairs of atoms, but $g_{AB}(r)$ and $g_{BA}(r)$ are identical, therefore in Fig. 2a we present only 3 functions.

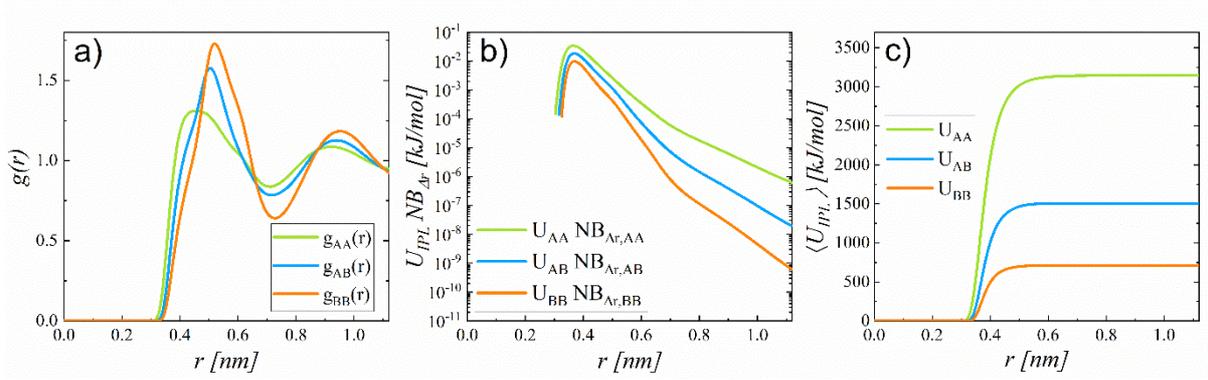

*Fig. 2*
*Panel a) depicts the pair-correlation function calculated between different types of atoms. Panel b) represents the contributions to the system's potential energy resulting from the neighbors of selected atom and separated by given distance. Panel c) shows how the potential energy of the system depends on the distance between interacting atoms.*

It can be seen that $A$-type atoms are able to approach each other at the closest distances, which is due to the least steep potential shape ($m_{AA} = 12$) observed at small intermolecular distances. On the other hand, the most narrow and prominent first peak is observed for atoms of type $B$, $m_{BB} = 18$. However, the most important feature of $g_{XY}(r)$ is that it enables the calculation of the average number of the $Y$-type atoms, which are the neighbors of $X$-type atom. If we consider the sphere of radius $R$, the number of neighbors is as follows, $NB_{XY} = \rho_Y \int_0^R 4\pi r^2 g_{XY}(r) dr$, $\rho_Y$ is the number density of $Y$-type of atoms. Hence, integrating $g_{XY}(r)$ over intervals, $\Delta r$, we get an average number of $Y$-type atoms that are separated from the reference $X$-type atom by a given distance, $r$. Those atoms, $NB_{XY,\Delta r}(r)$, contribute to the average potential energy according to the interaction potential, $U_{XY}(r)$. At this point it has to be mentioned that despite $g_{XY}(r) = g_{YX}(r)$, $NB_{XY,\Delta r}(r)$ might differ from $NB_{YX,\Delta r}(r)$ due to different number densities, $\rho_Y \neq \rho_X$. In Fig. 2b, we show the product of $NB_{\Delta r}(r)$ and $U_{IPL}(r)$ to analyze the contributions to the potential energies resulting from the neighbors located around on the given distance. Similar to the results presented in Fig. 2a, $NB_{XY,\Delta r}(r)U_{XY}(r)$ for mixed interactions are identical, thus, we present only one of them. As one can see, the discussed contributions exhibit an evident peak for all interactions. The maxima of those peaks are observed at similar $r$, which is due to identical $\sigma$ for all potentials. It is also worth noting that the results are presented in the logarithmic scale because $NB_{\Delta r}(r)U_{IPL}(r)$ significantly decreases when $r$ increases. It immediately suggests that contributions to $\langle U_{IPL} \rangle$ resulting from atoms separated by relatively long distances are irrelevant. However, to calculate the total $\langle U_{IPL} \rangle$ for many-atomic molecules, we have to consider the number of atoms belonging to the central molecule. In studied case, the RM is composed of 2 atoms of type A and 2 atoms of type B. Thus, each interaction occurs twice, $N_{at,AA} = N_{at,AB} = N_{at,BA} = N_{at,BB} = 2$. Each interaction has to be taken into account only once, thus the factor $N/2$ should be included. Finally, we obtain:

$$\langle U_{IPL} \rangle = \frac{N}{2} N_{at} \sum_{r=0}^{r=r_{cut}} NB_{\Delta r}(r) U_{IPL}(r). \qquad Eq.\ (6)$$

As it is presented in Fig. 2c, $\langle U_{IPL} \rangle$ becomes practically constant when $r$ is longer than $0.5\ nm$. It confirms that interactions between substantially separated molecules are not crucial for $\gamma$ value. Interestingly, $0.5\ nm$ does not seem to be a relatively long distance for the studied system because around $0.5\ nm$ the first peaks of $g(r)$ are observed. This observation immediately explains the literature reports suggesting that exclusively interactions between the nearest atoms are responsible for $\gamma$ value. [17,44] Combining Eq. ( 5) and Eq. ( 6), we calculate $\gamma_g = 4.64$, which is very close to that predicted by the isomorph theory and leads to accurate density scaling, results not presented.

As a final step, we test our interpretation of $\gamma$ and the method for estimating its value. Thus, we design a system, which is even more complex. The new RM is constructed from three different types of atoms, i.e., it consists of 2 atoms of type $A$ and 1 of type $B$ and $C$. See the inset in Fig. 3a where a scheme of the new RM is presented.

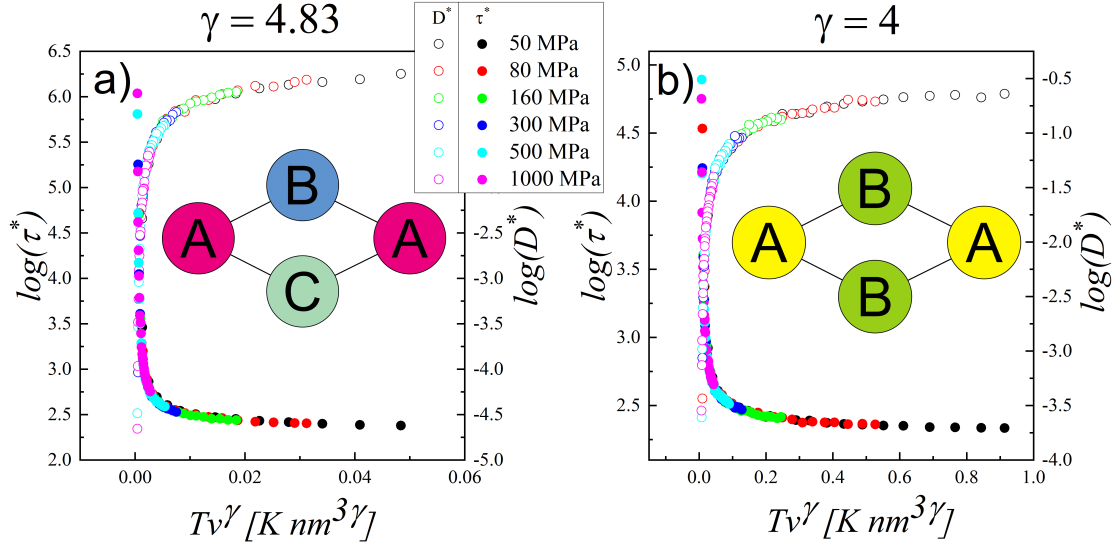

*Fig. 3*
*The density scaling of RM molecules constructed from 3 different atoms, which interaction potentials are described by various exponents of the repulsive term a) and 2 different atoms of the same exponent of the repulsive term but the different pre-power factors, b). The schemes of molecules are presented as an inset. The density scaling values are obtained using* Eq. (5) *and* Eq. (6)

Due to such architecture, we can distinguish six different interatomic interactions: $m_{AA} = 21$, $m_{BB} = 15$, $m_{CC} = 9$, $m_{BC} = 12$, $m_{AB} = 13$ and $m_{AC} = 14$. The experiment's procedure was exactly the same as for the first RM system. The *NVT* simulation had been preceded by simulation at *NPT* carried out at a wide range of temperatures and pressures. The determined $\tau^*$ and $D^*$ can be scaled according to Eq. (1), see Fig. 3a. Importantly, the used $\gamma$ value is estimated employing Eq. (5), Eq. (6) and $g(r)$ functions calculated at only one thermodynamic state, i.e., $T = 110\ K$ and $p = 300\ MPa$. This result confirms that the method proposed by us works for molecules composed of many atoms. Consequently, there is no obstacles to apply it for real systems. In this context, it is worth putting readers attention to the fact that the number density of molecules, which is used to estimate $\langle U_{IPL} \rangle$, is a constant at a given thermodynamic state. Hence, according to Eq. (5) it can be canceled, which implies that only a chemical composition of the molecule, and structure of the liquid are required to apply the proposed methods. The potentials' parameters are already defined and included in force fields.

At the end, we would like to test the prediction of Eq. (5), according to which $\gamma$ does not depend on $C$ parameter of the $IPL$ potential when the all interatomic interactions are parameterized using the same $m$ value. Consequently, we examined the RM system, which is composed of two types of atoms, where $C_{AA} = 2C_{BB} = \sqrt{2}C_{AB}$. The $\varepsilon$ and $\sigma$ are identical to those used in in our previous papers, whereas $m = 12$. The scheme of the molecule is presented in Fig. 3b. In this case we also apply the afore-described procedure of the experiment. As we present in Fig. 3b, the results for the studied system accurately scale with $\gamma = 4.0$, which is obtained by combining Eq. (5), Eq. (6), and $g(r)$ calculated at $T = 110\ K$ and $p = 300\ MPa$. The identical value of $\gamma$ was also reported by us in Ref. [41] for the analogic RM system characterized by the same $C$ value for all atoms. Hence, we can state that the tested prediction is valid.

Summarizing, in this article, we deliver the exact interpretation of the density scaling exponent value for van der Waals systems composed of polyatomic molecules, which atoms interact by purely repulsive potentials. In this case, $\gamma$ is a weighted average of the exponents describing particular intermolecular interactions, with weights determined by average potential energies originating from those interactions. Moreover, we prove that interactions between only the nearest atoms notably influence the density scaling and hence the dynamics, thermodynamics, and structure of the systems. Interestingly, following our idea, the well-known structure characteristic, i.e., pair correlation function, can be used to estimate the $\gamma$ value. Notably, it is sufficient to determine this characteristic at only one thermodynamic state. Consequently, we believe that a comprehensive understanding of the origin of the density scaling exponent is at our fingertips.

ACKNOWLEDGMENT

The authors are deeply grateful for the financial support by the Polish National Science Centre within the framework of the Maestro10 project (Grant No. UMO2018/30/A/ST3/00323).

AUTHOR DECLARATIONS

Conflict of Interest

The authors have no conflicts to disclose.

Author Contributions

**F. Kaśkosz**: Conceptualization, Investigation, Formal analysis, Visualization, Writing – original draft. **K. Koperwas**: Supervision, Writing – review & editing. **A. Grzybowski**: Supervision, Discussion. **M. Paluch**: Supervision, Discussion, Funding acquisition.

Data availability

Data will be made available on request.